\documentclass[bib]{ba}

\usepackage{epsfig}
\usepackage{bayes}

\usepackage{amsmath,amsfonts,amsthm,amssymb,latexsym}
\usepackage{natbib}

\newenvironment{kvothe}{\small\begin{quote}\begin{em}}{\end{em}\end{quote}\normalsize\vglue -0.2truecm}
\newcommand\dropcap\noindent

\begin{document}

\inserttype{article}
\author{Robert, C.P.}{%
  {\sc Christian P.~Robert}\\Universit\'e Paris-Dauphine, CEREMADE, and CREST, Paris
}	
\title[Reading Evidence and Evolution]{{\it{\bfseries Evidence and Evolution}\/}:\\ A review}

\maketitle

\begin{abstract}
{\em Evidence and Evolution: the Logic behind the Science} was published in \citeyear{sober:2008} 
by Elliott Sober. It examines the philosophical foundations of the statistical arguments used to evaluate
hypotheses in evolutionary biology, based on simple examples and likelihood ratios. The difficulty with
reading the book from a statistician's perspective is the reluctance of the author to engage into model
building and even less into parameter estimation. The first chapter nonetheless constitutes a splendid
coverage of the most common statistical approaches to testing and model comparison, even though the advocation
of the Akaike information criterion against Bayesian alternatives is rather forceful.
The book also covers an examination of the ``intelligent design" arguments against the Darwinian evolution 
theory, predictably if unnecessarily resorting to Popperian 
arguments to correctly argue that the creationist perspective fails to predict anything. The following chapters cover
the more relevant issues of assessing selection versus drift and of testing for the presence of a common ancestor.
While remaining a philosophy treatise, {\em Evidence and Evolution}
is written in a way that is accessible to laymen, if rather unusual from a statistician viewpoint, and
the insight about testing issues gained from {\em Evidence and Evolution} makes it a worthwhile read. 
\end{abstract}


\section{Introduction}
\newcommand\EE{{\em Evidence and Evolution}}
The book {\em Evidence and Evolution} is written by Elliott Sober, a philosopher of science who has worked on the notion of evidence, 
in the statistical meaning \citep{jeffreys:1939} of the word. All chapters are related to papers authored or co-authored by him
over the past years. {\em Evidence and Evolution} does not aim at demonstrating the validity of one
evolution theory against another but rather at validating statistical ways of testing such hypotheses. While Sober establishes
that creationism cannot be analysed within this framework because it fails to make predictions, the remaining chapters consider
ways to evaluate evidence about natural selection (against the drift alternative) and about common ancestry, without concluding
about those. Sober also relates very much to the original works of Charles Darwin, sometimes making too much of rather general
sentences of his, but this historical touch adds nonetheless to the already considerable appeal of the book.

First, an acknowledgement of my limitations is in order:
As a statistician, I cannot evaluate the philosophical relevance of the book, even though the arguments are quite accessible to a layman
like me. I however appreciate the critical assessment of Popper's testability criterion when applied to creationism as well as the
extensive coverage of statistical principles for testing. A philosophical perspective on \EE~ is given by \cite{pfeifer:2009}.
Furthermore, being equally a layman in population genetics and evolution biology, I have difficulties in assessing the impact of the 
debate about the construction of tests on biologists, as the examples seemed to be too formalised and simplistic to be realistic. My
review is therefore necessarily biased towards a statistician's perspective and hence maybe unnecessarily critical in terms of what I
perceive as a lack of proper modelling. Indeed, I bemoan the absence throughout the book of a genuine statistical framework that would
allow for a complete statistical analysis of even one single real dataset, including the estimation aspects that are bypassed in \EE, 
as this would illustrate much more clearly the concepts at work. 

In addition, while I understand the historical and philosophical appeals of discussing creationism, since Darwin was subjected to many 
attacks on this very issue, I am quite skeptical on the impact the book could have on the current debate. Unsurprisingly (and I will 
explain why below),
the book remains at a very general and hence vague level when discussing creationism. There are so many possible versions about the
intervention of a god or of another supernatural being in the management of the World that to pick one particular version would be like
grabbing at water, all other remaining versions emerging unscathed from a detailed criticism of the chosen version. But to maintain that
creationism is testable (in Popper's sense), as Sober does, is to open a similar Pandora box about which version he is considering. I thus
personally deplore the inclusion of this chapter in {\em Evidence and Evolution}, even though this position is not taken on statistical 
grounds.

The review linearly proceeds through the four chapters of the book, ``Evidence" (Chap.~1), ``Intelligent design" (Chap.~2),
``Natural selection" (Chap.~3), ``Common ancestry" (Chap.~4). Once again, it primarily focus on the statistical aspects of
the debate initiated by Sober without discussing the biological or philosophical consequences. Although a hard read at times,
proceeding through \EE~was a worthy and rewarding experience for me that led me to re-think the terms and objects of statistical
testing when applied to a scientific theory. The book is accessible to the layman (in any of those three fields) and I thus encourage
readers to progress through \EE.

\section{Statistical evidence}
\begin{kvothe}``Although frequency data and a well-supported empirical theory can provide a basis for
assigning prior probabilities, the principle of indifference cannot." {\em Elliott Sober, {\em Evolution and Evidence},}  p.27\end{kvothe}
The first chapter is fairly well written and presents a reasonable picture on the different perspectives 
(Bayesian, likelihood, frequentist) used for hypothesis testing and model choice, if missing references to
the relevant literature (for instance, \citealp{berger:wolpert:1988} is omitted when the likelihood principle
is discussed). Akaike's information criterion 
is promoted as {\em the} method of choice, but this is a well-established model choice tool that can be accepted at a 
general level. I paradoxically find the introduction to Bayesian principles to be overly long (as often the case
in cognitive sciences) since, as the author acknowledges from the start, ``Bayes Theorem is a result in mathematics
[that] is derivable from the axioms of probability theory" (page 8). This is especially blatant when considering that Sober
takes a very long while to introduce prior densities on parameter spaces, a reluctance that agrees with the parameter
free preferences of the book. The (standard) criticisms he addresses to the choice of those priors (which should be 
``empirically well-grounded", pages 26 and 27, as also pointed out in the above quote) 
will periodically resurface throughout the book, but they are far from convincing
as they mistake the role of the prior distributions as reference measures \citep{bernardo:smith:1994} for expressions of
truth. The extended criticism of the foundations of Neymann--Pearson testing procedures is quite thorough and could benefit
genuine statistician readers as well as philosophers and biologists. 

\begin{kvothe}``The Akaike framework makes plausible a mixed philosophy: instrumentalism for models, realism for fitted models" 
{\em Elliott Sober, {\em Evolution and Evidence},}  p.98\end{kvothe}
At a general level, I have two statistical difficulties with this chapter. First, while Sober introduces the Akaike information
criterion as a natural penalty for comparing models with different difficulties, I fear the notions of statistical parsimony and
dimension penalty are mentioned much too late in the chapter. Using a likelihood ratio for embedded models is for instance meaningless
unless a correction for the difference in dimensions is introduced. Second, the models used in this chapter and throughout the book are
singularly missing variable parameters, which makes all tests appear as comparison of point null hypotheses. The presence of nuisance
or interest parameters should be better acknowledged.

\begin{kvothe}``Bayesianism is a substantive epistemology, not a truism."
{\em Elliott Sober, {\em Evolution and Evidence},}  p.107\end{kvothe}
At a specific level, it is not possible to address all minor points with which I disagree but I think Sober is misrepresenting the
Bayesian approach to model choice and that he is missing the central role played by the Bayes factor in this approach. The fact that
the Bayes factor is an automated Ockham's razor with the proper penalty for differences in dimension \citep{mackay:2002} is altogether missed. 
In particular, Sober reproduces the error found in Templeton's (\citeyear{templeton:2010}). He indeed states that 
``the simpler model cannot have the higher prior probability, a point that Popper (1959) emphasised" (page 83).
And he insists further that there is no reason for thinking that
$$
P(\theta=0) > P(\theta>0)
$$
is true (page 84). (This common sense constraint obviously does not make sense for continuous state spaces since comparing models
requires working with foreign dominating measures.) Even though the likelihood ratio is a central quantity in the chapters to follow,
I am also reluctant to agree with introducing a specific category for likelihoodists (sic!) since, besides a Bayesian incorporation,
a calibrated likelihood leads either
to a frequentist Neymann--Pearson test or to a predictive tool like Akaike's (which is also frequentist in that it is an unbiased estimator).
In addition, the defence of the Akaike criterion is overdone, in particular the discussion about the unbiasedness of AIC which confuses
the fact that the averaged log-likelihood is an unbiased estimator of the Kullback--Leibler divergence with the issue that the AIC involves
a plug-in estimator of the parameters, as shown on pages 85 and 101. The arguments for AIC versus BIC are rather weak, from BIC being biased 
(correct but irrelevant) and Bayesian (incorrect), to the fact that it contradicts the above fallacious ordering of simple versus complex
models. A discussion of the encompassing framework of \cite{george:foster:2000} would have been most welcomed at this stage.

While the above points are due criticisms (from a statistician), the fact remains that this chapter is an exceptionally 
good and lucid discussion on the philosophy of testing and that it could well serve as the basis of a graduate reading
seminar. I thus recommend the to all statistician readers and teachers.

\section{"Intelligent design"}
    \begin{kvothe}``When dealing with natural things we will, then, never derive any explanations from the purpose which God 
or nature may have had in view when creating them and we shall entirely banish from our philosophy the search for final 
causes. For we should not be so arrogant as to suppose that we can share in God's plans." {\em Ren\'e Descartes, {\em Les 
Principes de la Philosophie},} Livre I, 28.\end{kvothe}
The second chapter of {\em Evidence and Evolution} tries to address the case of ``intelligent" design from an epistemological perspective,
namely as to whether or not evidence based reasoning can be applied to this line of thoughts. As pointed above, I was loath to get into 
this chapter for fear of being dragged into a barren and useless debate, but it stands the test from an academic perspective in that
the chapter is written from a purely philosophical perspective. I was originally expecting more statistical arguments given the tenor 
of the first chapter, but I realise this would have been a fruitless exercise given the infinite polymorphism of ``intelligent" design 
theories.\footnote{At a purely anecdotal level,
this chapter brought to me the interesting side informations that Jonathan Swift's {\em Gulliver's Travels} mentions a random word generator and
that John Arbuthnot created the John Bull character.}

\begin{kvothe}``Tout \'etant fait pour une fin, tout est n\'ecessairement pour la meilleure fin. Remarquez bien que les nez ont \'et\'e faits pour porter des lunettes, aussi avons-nous des lunettes." {\em Voltaire, {\em Candide},} Chapitre 1.\end{kvothe}
I find the introduction of the chapter interesting in that it relates the creationist thesis to a long philosophical tradition 
(witness the above quote from Descartes) rather than to the current unscientific debate about ``teaching" creationism in US and UK schools. 
The disputation of former theses like the one of Paley's watch is however taking most of the chapter and this is disappointing in my 
opinion.\footnote{The remark that ``Paley was well-aware of the relevant facts about monkeys and typewriters" (page 120) sounds like an
anachronism, since the typewriter still had to emerge in the early 1800's. A reference to Voltaire's {\em Candide} ridiculing design would
have been more appropriate.}
Somehow predictively (in the sense that this would have been my first argument), 
Sober mostly states the obvious when arguing that when gods or other supernatural beings enter the picture, 
they can explain for any observed fact with the highest likelihood while being unable to predict any fact not yet observed. 
I would have preferred to see hard scientific facts and the use of statistical evidence, even of the AIC sort, although as noted
previously I can see the ultimate lack of purpose in picking a specific version of creationism that could bend itself to a full
statistical analysis. While I consider Sober's analysis of Popper's testability to be of independent interest, 
as it rightly remarks that all {\em probabilistic theories} are unfalsifiable in Popper's sense (page 130), it does not bring further 
arguments because Sober coherently argues that even the theory of ``intelligent" design is falsifiable. 

    \begin{kvothe}``Bayesian philosophers of science see each hypothesis as competing with its own negation." {\em Elliott Sober, {\em Evolution and Evidence},} p.354\end{kvothe}
In Section 2.19 about model selection, the comparison between a single parameter model and a one million parameter model 
hints at Ockham's razor \citep{adams:1987,berger:jefferys:1992,sober:1994,mackay:2002}, but, once again,
Sober misses the point about a major aspect of Bayesian analysis. Indeed, through the use of hyperpriors and 
hyperparameters, observations about one group of parameters also bring information about other groups of parameters when those are 
related via a hyperprior (as in small area estimation). Given that the author hardly ever discusses the use of priors over the model 
parameters and seems to instead rely plug-in estimates, he does not take advantage of the marginal posterior dependence between the 
different groups of parameters.

\section{Testing for selection}
    \begin{kvothe}``To test a theory, you need to test it against alternatives." {\em Elliott Sober, {\em Evolution and Evidence},}  p.190\end{kvothe}
The ``Natural selection" chapter is difficult to read for a layman like me in that it seems overly repetitive, 
using somehow obvious arguments while missing clearcut conclusions and directions. This bend must be due to the philosophical 
priorities of the author but, despite opposing Brownian motion to Ornstein-Uhlenbeck processes at the beginning of the chapter 
(which would have made for a neat parametric model comparison setting),\footnote{Both processes are dubbed {\em natural} by the author
for pure drift and selection plus drift, respectively (page 194). I would thus avoid this qualification since these are mathematical
models and not natural phenomena. Similarly, mentioning a {\em bell curve} on page 197 gives a confusing message since the
setting is not necessarily linked with the normal distribution.} there is no quantitative argument nor illustration found 
in this third chapter that would relate to statistics. This is unfortunate as the questions of interest (testing for natural 
selection versus pure drift or versus phylogenetic inertia or yet for tree structure in phylogenetics) could clearly be conducted at a 
numerical level as well, through the AIC factor or through a Bayesian alternative. The aspects I found most interesting in this chapter 
may therefore be deemed as marginalia by most readers, namely (a) the discussion whether or not the outcome of a test should at all depend on 
the modelling assumptions (the author seems to doubt this, hence relegating Bayesian techniques to their dust-gathering shelves), 
and (b) the point that parsimony is not a criterion per se.

    \begin{kvothe}``What we need is a probability distribution of the different values $A$ might have, conditional on each hypothesis." {\em Elliott Sober, {\em Evolution and Evidence},}  p.210\end{kvothe}
About the first point, the philosophical stance of the author is not completely foolproof in that he concedes---witness the above quote---that 
testing hypotheses without accounting for the alternative is not acceptable. This would almost irremediably call for a truly Bayesian resolution,
but my impression is that Sober looks at the problem from a purely dichotomous perspective, either the 
hypothesis or the alternative being true. This is a bit of a caricatural representation as 
he incorporates the issue of calibrating parameters under 
the different hypotheses, and there is a sort of logical discrepancy lurking in the background of the argument. Again, working out a fully 
Bayesian analysis of a philogenic tree---mentioned on page 190 as part of the model assumptions---would have clarified the issue immensely. 
And rejecting Bayesianism on the ground that ``there is no objective basis for producing an answer" (page 239) is limited on the epistemological 
side. This is particularly frustrating when considering the above quote where Sober acknowledges the need for a posterior distribution over
the ancestral trait $A$ and where he advocates using ``equilibrium probabilities as priors for the state of the ancestor $A$" (page 211).

Even though I understand that the book 
is not trying to debate about the support for a specific evolutionary hypothesis but rather about the methods used to test such hypotheses 
and the logic behind these, completely worked-out example would have made my appreciation (and maybe other readers') of Sober's points much 
easier. To mention one such issue, the construct of an efficiency or fitness function (discussed throughout the chapter, see, 
e.g., page 196) that could drive the natural selection is not discussed from a realistic biological perspective but strikes me 
instead as a purely formal entity. (The exception being the aphid eating time of Figure 3.8.)
Note that Section 3.9 is more model oriented, using molecular data, although neither the {\em significantly different} (page 238)
results nor the Akaike score are given. (There is also a conceptual mistake there in that the neutral hypothesis is stated as $d_{13}-d_{23}=0$
instead of $E[d_{13}-d_{23}]=0$, see Figure 3.25 for a similar confusion.)
Thus, I fail to see who could take benefit from reading this chapter as a whole---even though particular points are worthwhile
contributions to the philosophy of testing. For instance, a biologist will most likely process the arguments 
and illustrations provided by Sober but this biologist could leave the chapter with a feeling of frustration 
at the apparent lack of conclusion. (As a 
statistician, I fail to understand how the likelihoods repeatedly mentioned by Sober can be computed because they never involve any parameter.)

    \begin{kvothe}``Parsimony does not provide a justification for ignoring the data." {\em Elliott Sober, {\em Evolution and Evidence},}  p.250\end{kvothe}
Since I believe that the Ockham's razor argument has had a global negative impact on the understanding of the
parsimony requirement in testing \citep{mackay:2002,robert:2007}, I find the warning signals about parsimony 
(given in the last third of the chapter) more palatable. Parsimony being an ill-defined concept, especially from a statistical perspective 
(where even the dimension of the parameter space is debatable, \citealp{spiegbestcarl}), no model selection is acceptable 
if only based on this argument. Note further that parsimony is understood in two different ways in the book, one being 
connected to the Ockham's razor and one leading to the specific phylogenetic parsimonious reconstruction (defined on page 207 as the 
``minimizing the total amount of evolution that must have occurred in the genetic tree", although I fail to understand the numerical
illustration provided on the same page). In addition, \cite[][Chapter 7]{sober:1994} makes a similar point in somehow more accessible
(if non-statistical) terms.

\begin{kvothe}
    ``Instead of evaluating hypotheses in terms of how probable they say the data are, we evaluate them by estimating how accurately they'll predict new data when fitted to old." {\em Elliott Sober, {\em Evolution and Evidence},}  p.229\end{kvothe}
The chapter also addresses the distinction between hypothesis testing and model selection as paramount---a point I subscribed to for a long 
while before getting convinced of the opposite---, but I cannot get to the core of this argument. It seems 
Sober sees model selection through the predictive performances of the models under comparison, if the above quote is representative of his 
thesis. (Overall, I find the style of the chapter slightly uneven, as if the fact that some sections like Section 3.7 are adapted 
from earlier papers would make for different levels of depth.)

Statistically speaking, this chapter also has a difficulty with the continuity assumption. To make this point more precise, I notice there is 
a long discussion about reaching the optimum configuration (for polar bear fur length) under the SPD hypothesis, but I think evolution happens 
in discontinuous moves. (Think for instance of changes in the number of chromosomes.)
The case about the existence of a local minimum in Section 3.4 and the difficulty in moving from a local mode
with a global mode is characteristic of this difficulty as a ``valley" on a ``fitness curve" 
that in essence takes three possible values over the three different types of eye designs does not really constitute a bottleneck in the 
optimisation process. Similarly, the temporal structure of the statistical models in Sections 3.3 and 3.5 is never mentioned, even though 
it needs to be defined for the tests to take place. (For instance, in several places, time is mentioned without a clear definition. I have
trouble to understand how a finite time or even the original time $t=0$ can be assessed in such settings.)
The past versus current convergence to stationarity or equilibrium and hence to optimality 
under the SPD hypothesis is an issue and so is the definition of time in the very simple $2\times 2$ Markov chain example? 
And given a $2\times 2$ contingency table like
$$
\begin{matrix} &\text{fixed} &\text{polymorphic}\\ \text{synonymous} &17 &42 \\ \text{nonsynonymous} &7 &2\\ \end{matrix}
$$
testing for independence between both factors is a standard among the standards: I thus fail to understand the lengthy and 
inconclusive discussion of pages 240-243.\footnote{The presentation of the $2\times 2$ contingency tables in Figure 4.11 is fairly unusual in that
the position of the counts and the factor values are inverted.} Another statistical difficulty relates to the implicit use of plug-in estimates
and to Sober's reluctance to adopt Bayesian arguments based on marginals:
in the discussion on pages 255-256 about inferring on phylogenetic trees, the Markovian independence between branches of the trees
given the first level ancestors is confused with the impossibility of running inference ``on ancestors that are `deeper.'" The
independence between non-contiguous nodes of the tree only holds conditional on the intermediate nodes. It vanishes when integrating
over the intermediate nodes.

\section{Common or separate ancestry}
    \begin{kvothe}
``Darwinians would not be satisfied if all life on Earth derived from the same large slab of rock."  p.269\end{kvothe}
The final chapter of the book (apart from the concluding summary) is about common ancestry and may be the most statistically oriented of 
the three last chapters. This is not to say the chapter is without defaults, including in particular a certain tendency to repeat 
the same arguments. but this is somehow the chapter I appreciated the most. The chapter starts with a detailed analysis on how the hypothesis 
of common ancestry should be set, the main distinction being between one organism and several, while pointing out the confusing effect of 
lateral gene transfer.  Inference about phylogenetic trees and the use of genetic sequences rather than simplistic traits gets us closer 
to the true issues at stake. Another interesting feature of this chapter is the relation to Darwin's reflections on the common origin of 
life on Earth through many quotes.

    \begin{kvothe}
``If those prior probabilities are obscure, the same will be true of the posterior probabilities." {\em Elliott Sober, {\em Evolution and Evidence},}  p.277\end{kvothe}
The statistical issue is thus of testing for a common ancestor versus separate ancestors for a set of organisms. The nature of the information 
contained in the data is never made precise enough to understand whether this fits the principle of total evidence stressed throughout the book. 
The chapter also shows a more lenient disposition towards Bayesian solutions (relying on priors on page 301) 
but Section 4.3 ends up with a damning statement, due to the 
impossibility of defining an objective prior because Sober wants prior probabilities that have some authority. This is a self-defeating 
constraint leading to {\em empirically well-grounded priors} (page 276).

    \begin{kvothe}``Those propositions suffice for similarity to be evidence for common ancestry, and they have broad applicability." {\em Elliott Sober, {\em Evolution and Evidence},}  p.283\end{kvothe}
The part about Reichenbach's (1956) sufficient condition for a common trait to induce a likelihood ratio larger than one in favour of the 
common ancestor hypothesis needs to be discussed as this is the point I find the most puzzling in the chapter. Indeed, most of the 
nine assumptions of Reichenbach (1956) relate both models under comparison, i.e.~common ancestry versus separate ancestry by 
$$
\text{Pr}(X=i|Z=j) = \text{Pr}(X=i|Z_1=j)\,,\quad
\text{Pr}(Y=i|Z=j) = \text{Pr}(Y=i|Z_2=j)\,, 
$$ 
and
$$
\text{Pr}(Z=j) = \text{Pr}(Z_1=j) = \text{Pr}(Z_2=j)\,, 
$$
where $X$ and $Y$ are the observed character traits for two species, and where $Z$ is the common ancestor trait, while $Z_1$ and $Z_2$ are
the separate ancestor traits. This type of assumptions is statistically and philosophically meaningless in the sense that
models under comparison should not share any of their parameters. If the point is about determining which model is ``true", 
the ``wrong" model does not exist, there is no $Z$ or no $(Z_1,Z_2)$, hence 
the corresponding parameters do not have any substance either. For instance, when building a Bayesian model 
to compare those models, there exists a separate prior distribution on each group of parameters. The common parameter assumption is
thus not compatible with selecting one of the two models. This unrealistic framework may be the result of a reluctance to 
handle true (i.e., unknown) parameters as in a regular statistical 
analysis.  (See, e.g., the lament that ``until values for adjustable parameters are specified, we cannot talk about the probability of the data 
under different hypotheses", page 338.) What is striking is the reliance of the whole chapter on this unnatural set of hypotheses since it keeps 
resurfacing throughout the chapter. Sober writes that Propositions 1-9 are not consequences of the axioms of probability. Neither are they 
necessary conditions for common ancestry to have a higher likelihood than separate ancestry (page 283). Nonetheless, this is creating an
unnecessary bias in the perception of the problem which may induce critics of evolution to reject the whole approach.

    \begin{kvothe}``If there was no such common ancestor, what would alignment ever mean?" {\em Elliott Sober, {\em Evolution and Evidence},}  p.291\end{kvothe}
The theme of the missing model I have alluded to in the previous posts is also recurrent in this chapter. There are a lot of paragraphs about 
the choice of the representation of the difference between two species, from trait to gene sequence, and the author acknowledges that the 
difficulty in this choice has to do with a requirement for a more advanced theoretical representation (model) adapted to more complex data. 
This sounds rather obvious stated that way but the book wanders around this point for pages. (An example is the above quote that misses the 
point about sequence alignment: this is a perfectly well-defined measure of distance, common ancestor\footnote{An interesting discussion
appears on page 291 about the bias induced by alignment, the conclusion being that ``aligning sequences is not loading the dice". I think
that alignment is akin to maximum likelihood plug-in and that it is therefore favouring the null hypothesis. It should therefore be
accounted for in the statistical procedure.} or not.) And the overall conclusion is 
a vague call for the principle of total evidence (which is a rephrasing of the likelihood principle), after rightly dismissing the
majority rule (page 295). As illustrated by the section on multiple 
characters, the discussion is confusing without a proper model. It is only on page 300 of the book that a completely defined model for the 
evolution of a dichotomous trait (i.e. the simplest possible case) appears.\footnote{A minor point of contention there
is the use of {\em bias} for the Markovian model on chromosome transformation, where {\em reversibility} would have been more appropriate.} 
This model is a rather crude tool, as it depends on arbitrary calibration factors like $P(Z=0)=0.99$ (instead of the absorbing $1$) 
and, more importantly, on an unspecified time $t$ (as in ``what time is it on the evolution clock?"). The corresponding likelihood ratio is 
then (under one of the selection schemes)
$$
\dfrac{0.01b_t^2 + 0.99}{[0.01b_t+0.99]^2}
$$
where the dependence on those calibration factors is obvious. This illustrates the impossibility to reach a satisfactory conclusion without going 
first through a statistical analysis of the problem.

Although this is not the purpose of the book, I think the debate about causality found therein is rather superficial. For instance, 
while, in Section 3.6, causality and correlation are differentiated (see footnote 22 on page 224 and page 233) 
Section 3.8 embarks upon testing for a 
causal connection, discussing \cite{reichenbach:1956} without mentioning the Humean thesis of the logical and statistical impossibility
of such a test. Most of Chapter 4 is about testing ``whether there was a common {\em cause}" (page 247). A notion of information is 
mentioned on page 305 without being defined and I do not understand whether or not this relates to Fisher's information 
\citep{lehmann:casella:1998} or to the Kullback divergence, as, apparently, no parameter is involved.

    \begin{kvothe}``It is possible for data to discriminate among a set of hypotheses without saying anything about a proposition that is common to all the alternatives considered." {\em Elliott Sober, {\em Evolution and Evidence},}  p.315\end{kvothe}
The debate about the phylogenetic tree reconstruction versus the test for common ancestry (Sections 4.7 and 4.8) lacks appeals for 
the very reason exposed above. The tree structure may be incorporated within the model(s) and integrated out in a Bayesian fashion 
to provide the marginal likelihood of the model(s). Although this seems to be an important issue, as illustrated by the controversy 
with Templeton \citep{templeton:2008,clade:2010}, 
the opposition between likelihood inference and ``cladistic" parsimony is not properly conducted in that, as a na{\"\i}ve reader, 
I cannot understand Sober' presentation of the later. This section is much more open to Bayesian processing by abstaining from the 
usual criticism about the lack of objectivity of the prior selection, but it entirely misses the ability of the Bayesian approach 
to integrate out the nuisance parameters, whether they are the tree topology (standard marginalisation) or the model index (model 
averaging). The debate about the limited meaning of statistical consistency is making the valid point that consistency only puts light 
on the case when the hypothesised model is true, but extended consistency could have been considered as well, namely that the procedure 
will bring the hypothesised model as close as possible to the ``true" model within the hypothesised family of models. What I gather from 
this final section is that cladistic parsimony tries to do without models (if not without assumptions), which seems to relate to Templeton's 
views about Bayesian inference.

This is the most enjoyable chapter of the book from my point of view, even though the lack of real illustrations makes it less potent than 
it could be. It also shows the limitation of a philosophical debate on simplistic idealisations of the real model. The book rarely acknowledges 
(see pages 236 and 334) that genealogical hypotheses are composite. An incorporation of the parameter 
estimation in the inferential process would have improved the depth of the debate.

\section{Conclusion}

    \begin{kvothe}``A quantitative assessment of goodness of fit is indispensable when evolutionary models are compared." {\em Elliott Sober, {\em Evolution and Evidence},}  p.362\end{kvothe}
{\em Evidence and Evolution} is very well-written, with hardly any typo (the unbiasedness property of AIC is stated at the bottom of
page 101 with the expectation symbol $E$ on the wrong side of the equation, 
Figure 3.8c is used instead of Figure 3.7c on page 204, 
Figure 4.7 is used instead of Figure 4.8 on page 293, 
Simon Tavar\'e's name is always spelled Taver\'e,
{\em vaules} rather than {\em values} is repeated four times on page 339).
The style is sometimes too light and often too verbose, with an abundance of analogies that I regard as sidetracking, 
but this makes for an easier reading (except for the sentence ``the key 
to answering the second question is that the observation that $X=1$ and $Y=1$ produces stronger evidence favoring CA over SA the {\em lower}
the probability is that the ancestors postulated by the two hypotheses were in state 1", on page 314, that still eludes me!). As detailed in
this review, I have points of contentions with the philosophical views about testing in {\em Evidence and Evolution}
as well as about the methods exposed therein, but this does not detract from the appeal of reading the book. (The lack of completely
worked out statistical hypotheses in realistic settings remains the major issue in my criticism of the book.) While the criticisms of
the Bayesian paradigm are often shallow (like the one on page 97 ridiculing Bayesians drawing inference based on a single observation),
there is nothing fundamentally wrong with the statistical foundations of the book.
I therefore repeat my earlier recommendation in favour of {\em Evidence and Evolution}, Chapters 1 and (paradoxically) 5 being the easier
entries. Obviously, readers familiar with Sober's earlier papers and books will most likely find a huge overlap with those but others will
gather Sober's viewpoints on the notion of testing hypotheses in a (mostly) unified perspective.

\section*{Acknowledgements}
The author's research is partly supported by the Agence Nationale de la Recherche (ANR, 212,
rue de Bercy 75012 Paris) through the 2007--2010 grant ANR-07-BLAN-0237 ``SPBayes". 

\renewcommand{\bibsection}{\section*{References}}

\bibliographystyle{ba}

\end{document}